\documentclass[prd, 
showpacs,nofootinbib,preprintnumbers]{revtex4}
\usepackage{amsmath}
\usepackage{amssymb}
\usepackage{amsfonts}
\usepackage{graphicx}
\usepackage{dcolumn}
\usepackage{hyperref}

\begin{document}

\title{On the $\Lambda$CDM Universe in  $f(G)$ gravity}
\author{Ratbay Myrzakulov$^1$, Diego S\'aez-G\'omez$^2$ and Anca Tureanu$^3$}

\affiliation{$^1$Dept. Gen. Theor. Phys., Eurasian National University, Astana, 010008, Kazakhstan}
\affiliation{$^2$Institut de Ci\`{e}ncies de l'Espai ICE/CSIC-IEEC, Campus UAB, Facultat de Ci\`encies, TorreC5-Parell-2a pl,E-08193 Bellaterra (Barcelona) Spain}
\affiliation{$^3$Helsinki Institute of Physics, P.O. Box 64, FI-00014 Helsinki, Finland}

\begin{abstract}
In the context of the so-called Gauss-Bonnet gravity, where the 
gravitational action includes function of the Gauss-Bonnet invariant, 
we study cosmological solutions, especially the well-known 
$\Lambda$CDM model. It is shown that the dark energy contribution 
and even the inflationary epoch can be explained in the frame of 
this kind of theories with no need of any other kind of component. 
Other cosmological solutions are constructed and the rich properties 
that this kind of theories provide are explored.
\end{abstract}

\maketitle

\section{Introduction}

It is widely accepted by the scientific community that our Universe 
is currently in an accelerated phase. General Relativity in its 
standard form can not explain the accelerated expansion without 
extra terms or components, which have been gathered under the name 
of dark energy. Since the discovery of the acceleration, a large 
number of possible mechanisms have been proposed to explain the 
origin of the dark energy, from the cosmological constant, scalar 
fields, to modifications of general relativity as well as other 
alternatives (for reviews on unified inflation-dark energy modified gravities, see Refs. 
\cite{N1}-\cite{ReviewN_S}, and for comparison with observational data see Refs.~\cite{C1}-\cite{C2}). The most popular idea is represented by the cosmological 
constant, whose origin may be explained by means of the vacuum 
energy density, although its value does not match the one predicted 
by quantum field theories. Even in such a case, the Equation of 
State (EoS) parameter is a constant equal to $-1$, while some recent 
observations suggest that the EoS parameter is a dynamical variable 
and it could have crossed the phantom barrier, with the EoS 
parameter being less than  $-1$. Such a dynamical behavior of the 
EoS could be well explained by scalar fields with quintessence or 
phantom behavior (see Ref. \cite{Eli1}). Another alternative is the 
modification of the gravity law. Several ways have been suggested to 
perform such a modification of General Relativity by means of the 
modification of the Hilbert-Einstein action, which obviously 
modifies the field equations. In this framework, the so-called 
$F(R)$ gravity has been explored, which introduces a more complex 
function of the Ricci scalar in the action and explains well the 
cosmic history via cosmological reconstruction (see Refs. \cite{N7}-\cite{HLF(R)}).

In this work we study the so-called Gauss-Bonnet gravity, where the 
gravitational action includes functions of the Gauss-Bonnet 
invariant. This kind of theories have been investigated and may 
reproduce the cosmic history (see Refs. \cite{N2}-\cite{EMOS}). 
Here, we show that the $\Lambda$CDM model can be well explained with 
no need of a cosmological constant but with the inclusion of terms 
depending on the Gauss-Bonnet invariant in the action. Even more, it 
is shown that the extra terms in the action coming from the 
modification of gravity could behave relaxing the vacuum energy 
density, represented by a cosmological constant, and may resolve the 
so-called cosmological constant problem.  To study and reconstruct 
the theory that reproduces such a model as well as other kind of 
solutions studied, we shall use a method proposed in Ref. \cite{N4} 
for $f(R)$ gravity and implemented for $f(G)$ gravity in Ref. 
\cite{EMOS}, where the FLRW equations are written as functions of 
the so-called number of e-foldings instead of the cosmic time. The 
possible phantom epoch produced by this kind of theories is also 
explored, as well as other interesting cosmological solutions, where 
the inclusion of other contributions as perfect fluids with 
inhomogeneous EoS are studied.

\section{$[R+f(G)]$ gravity}

We consider the following action, which describes General Relativity 
plus a function of the Gauss-Bonnet term (see Refs.~\cite{N2} and 
\cite{N9}):
\begin {equation}
S=\int 
d^{4}x\sqrt{-g}\left[\frac{1}{2\kappa^{2}}R+f(G)+L_{m}\right]\,, 
\label{1.1}
\end{equation}
where $\kappa^2=8\pi G_N$, $G_N$ being the Newton constant, and the 
Gauss-Bonnet invariant is defined as usual:
\begin{equation}
 G=R^2-4R_{\mu\nu}R^{\mu\nu}+R_{\mu\nu\lambda\sigma}R^{\mu\nu\lambda\sigma}\ .
\label{I2}
\end{equation}
By varying the action over $g_{\mu\nu}$, the following field equations 
are obtained:
$$
0=\frac{1}{2k^2}(-R^{\mu\nu}+\frac{1}{2}g^{\mu\nu}R)+T^{\mu\nu}+\frac{1}{2}g^{\mu\nu}f(G)-2f_{G}RR^{\mu\nu}+4f_{G}R^{\mu}_{\rho}R^{\nu\rho}
$$
$$
-2f_{G}R^{\mu\rho\sigma\tau}R^{\nu}_{\rho\sigma\tau}-4f_{G}R^{\mu\rho\sigma\nu}R_{\rho\sigma}+2(\nabla^{\mu}\nabla^{\nu}f_{G})R
-2g^{\mu\nu}(\nabla^{2}f_{G})R-4(\nabla_{\rho}\nabla^{\mu}f_{G})R^{\nu\rho}
$$
\begin {equation}
-4(\nabla_{\rho}\nabla^{\nu}f_{G})R^{\mu\rho}+4(\nabla^{2}f_{G})R^{\mu\nu}+4g^{\mu\nu}(\nabla_{\rho}\nabla_{\sigma}f_{G})R^{\rho\sigma}
-4(\nabla_{\rho}\nabla_{\sigma}f_{G})R^{\mu\rho\nu\sigma}\,,
\label{1.2}
\end{equation}
where we made the notations $f_G=f'(G)$ and $f_{GG}=f''(G)$. We shall assume throughout the paper a spatially-flat FLRW universe,  whose  metric is given by 
\begin {equation}
ds^{2}=-dt^{2}+a(t)^{2}\sum^{3}_{i=1}(dx^{i})^{2},
\label{FRW}
\end{equation}
where $a(t)$ is the scale factor at cosmological time $t$. For the  
metric (\ref{FRW}), the field equations give the FLRW equations, 
with the form
\[
0=-\frac{3}{\kappa^{2}}H^{2}+G 
f_{G}-f(G)-24\dot{G}H^{3}f_{GG}+\rho_{m}\,,
\]
\begin {equation}
0=8H^2\ddot{f}_{G}+16H(\dot{H}+H^2)\dot{f}_G+\frac{1}{\kappa^2}(2\dot{H}+3H^2)+f-Gf_G+p_m\,.
\label{1.3}
\end{equation}

The Hubble rate $H$ is here defined by $H=\dot{a}/a$, while the 
matter energy density $\rho_{m}$ satisfies the standard continuity 
equation:
\begin {equation}
\dot{\rho_{m}}+3H(1+w)\rho_{m}=0\ , \label{1.5}
\end{equation}
while  the Gauss-Bonnet invariant $G$ and  
the Ricci scalar $R$ can be defined as functions of the Hubble parameter as
\begin {equation}
G=24(\dot{H}H^{2}+H^{4}),\quad R=6(\dot{H}+2H^{2}).
\end{equation}
Let us  now rewrite Eq.~(\ref{1.3}) by using a new variable,
$N=\ln\frac{a}{a_{0}}=-\ln(1+z)$, i.e. the number of e-foldings, 
instead of the cosmological time $t$, where  $z$ is the redshift 
(this method has been implemented in Ref.~\cite{N4} for $f(R)$ 
gravity). The following expressions are then easily obtained
\begin{equation}
a=a_{0}e^{N},\quad H=\dot{N}=\frac{dN}{dt}\ , \quad
\frac{d}{dt}=H\frac{d}{dN}\ , \quad
\frac{d^2}{dt^2}=H^2\frac{d^2}{dN^2}+HH^\prime\frac{d}{dN}\ ,\quad
H^{\prime}=\frac{dH}{dN}\ .
\label{1.6}
\end{equation}
Eq.~(\ref{1.3}) can thus be expressed as follows
\begin{equation}
0=-\frac{3}{\kappa^{2}}H^2+24H^{3}(H'+H)f_{G}-f-576H^{6}\left(HH''
+3H'^{2}+4HH'\right)f_{GG}+\rho_{m}\ ,
\label{1.7}
\end{equation}
where $G$ and $R$ are now
\begin {equation}
G=24(H^{3}H'+H^{4})\ ,
\quad\dot{G}=24(H^{4}H''+3H^{3}H'^{2}+4H^{4}H')\ , \quad R=
6(HH'+2H^{2})\ .
\label{1.8}
\end{equation}
By introducing a new function $x$ as $x=H^2$, we have
\begin {equation}
H'=\frac{1}{2}x^{-1/2}x'\ ,\quad
H''=-\frac{1}{4}x^{-3/2}x'^{2}+\frac{1}{2}x^{-1/2}x''\ .\end{equation}
Hence,  Eq.~(\ref{1.7})  takes the form
\begin{equation}
0=-\frac{3}{\kappa^{2}}x+12x(x'+2x)f_{G}-f-24^{2}x\left[\frac{1}{2}x^{2}x''
+\frac{1}{2}xx'^{2}+2x^{2}x'\right]f_{GG}+\rho_{m}\ ,
\label{1.9}
\end{equation}
where we have used the expressions
\begin{equation} G=12xx'+24x^{2},\quad
\dot{G}=12x^{-1/2}[x^2x''+xx'^2+4x^{2}x']\  \quad \text{and} \quad 
R=3x'+12x. \label{1.10}
\end{equation}
Then, by using the above reconstruction method, any cosmological 
solution can be achieved, by introducing the given Hubble parameter 
in the FRW equations, which leads to the corresponding Gauss-Bonnet 
action.

\section{Reconstructing $\Lambda$CDM model in $R+f(G)$ gravity}

We are now interested to reconstruct $\Lambda$CDM solution in 
$R+f(G)$ gravity for different kind of matter contributions. It is 
well known that such a solution can be achieved in GR by introducing 
a cosmological constant (cc) in the action. Nevertheless, we show 
that in GB gravity there is no need of a cc. The cosmological models 
coming from the different versions of modified GB gravity considered 
will be carefully investigated with the help of several particular 
examples where calculations can be carried out explicitly.

In this paper,  we restrict ourself to explore some "classical" 
modified gravity models for the $\Lambda$CDM case, as well as other 
interesting and important cosmological solutions. For the 
$\Lambda$CDM model, the Hubble rate is given by
\begin{equation}
H^{2}= \frac{\Lambda}{3}+\frac{\rho_0}{3a^3},
\label{LambdaCDM}
\end{equation}
where $\rho_0$ is the matter density (which consists of barionic 
matter and cold dark matter) and $\Lambda$ is the cosmological 
constant. In the rest of this section we put $\kappa^2=1$. 

 For the  
$\Lambda$CDM model, described by the Hubble parameter 
(\ref{LambdaCDM}), we can write the derivatives of the scale factor 
as well as the Hubble parameter in the following  useful way:
\[
\dot{a}=\sqrt{\frac{\Lambda a^2}{3}+\frac{\rho_0}{3a}}\ , \quad \ddot{a}=\frac{2\Lambda a^3-\rho_0}{6a^2}\ ,
\]
\begin{equation}
\dot{H}=-\frac{\rho_0}{2a^3}=\frac{3}{2}\left(\frac{\Lambda}{3}-H^{2}\right)\ 
, \quad \ddot{H}=\frac{3\rho_0}{2a^3}\sqrt{\frac{\Lambda 
}{3}+\frac{\rho_0}{3a^3}}=\frac{9}{2}\left(H^{2}- 
\frac{\Lambda}{3}\right)H\ .
\end{equation}
Using these formulas we get
\[
R=4\Lambda+\frac{\rho_0}{a^3}\ , \quad 
G=24\left(\frac{\rho_0}{3a^3}+\frac{\Lambda 
}{3}\right)\left(\frac{\Lambda }{3}-\frac{\rho_0}{6a^3}\right)\ ,
\]
\begin{equation}
\dot{R}=\frac{-3\rho_0}{a^3}\sqrt{\frac{\rho_0}{3a^3}+\frac{\Lambda 
}{3}}\ , \quad 
\dot{G}=\frac{4\rho_0}{a^3}\left(\frac{2\rho_0}{a^3}-\Lambda\right)\sqrt{\frac{\rho_0}{3a^3}+\frac{\Lambda 
}{3}}\ .
\end{equation}
Then, the following relation  between $R$ and $G$ holds:
\begin{equation}
G=-\frac{4}{3}(R^2-9\Lambda R+18 \Lambda^2)\ .
\end{equation}
Let us recall that $x=H^2$. Then, some of the above formulas take the 
form:
\[
\dot{H}=\frac{1}{2}(\Lambda-3x)\ \quad
\ddot{H}=\frac{3}{2}(3x- \Lambda)\sqrt{x}\ , \quad R=3(\Lambda+x)\ ,
\]
\begin{equation}
G=12x(\Lambda -x)\ , \quad \dot{R}=3(\Lambda-3x)\sqrt{x}\ , \quad
\dot{G}=12(3x-\Lambda)(2x-\Lambda)\sqrt{x}\ .
\end{equation}
Note that  the  variable $x$ can be expressed in terms of $R$ or  
$G$ as
\begin{equation}
x=\frac{R}{3}-\Lambda \quad \text{or} \quad 
x=\frac{3\Lambda\pm\sqrt{9\Lambda^2-3G}}{6}\,,
\end{equation}
respectively. The above formulas will be useful to reconstruct the 
$\Lambda$CDM model as well as other cosmological solutions in the 
context of Gauss-Bonnet gravity, as it is shown below.

We write the first Friedmann equation (\ref{1.3}) in  the form
\begin {equation}
0=-3H^{2}+12H^2(\Lambda -H^2) 
f_{G}-f-288H^4(3H^2-\Lambda)(2H^2-\Lambda)f_{GG}+\rho_{m}\,,
\end{equation}
or
\begin {equation}
0=-3x+12x(\Lambda -x) 
f_{G}-f-288x^2(3x-\Lambda)(2x-\Lambda)f_{GG}+\rho_{m}\,.
\end{equation}
For further algebra more  convenient is the following form of  this  equation 
\begin {equation}
0=(\rho_{m}-3x-f)(\Lambda-2x)+[48x^2(3x-\Lambda)+x(\Lambda-x)]f_x+24x^2(3x-\Lambda)(\Lambda-2x)f_{xx}\ .
\label{ee}
\end{equation}
Now we wish to construct some particular exact solutions of this 
equation.

\subsection{Case I: $\rho_m=0$}

First of all, let us  consider the simple case in absence of matter, $\rho_m=0$. Then, the equation (\ref{ee}) takes the form
\begin {equation}
0=-(3x+f)(\Lambda-2x)+[48x^2(3x-\Lambda)+x(\Lambda-x)]f_x+24x^2(3x-\Lambda)(\Lambda-2x)f_{xx}\ .
\label{C1}
\end{equation}
We can analyze the cases where the cosmological constant term in the 
solution (\ref{LambdaCDM}) vanishes and where it is non-zero.
\begin{itemize}
    \item Let $\Lambda=0$. Then Eq. (\ref{C1}) reads as
\begin {equation}
0=-2(3x+f)-x(144x-1)f_x+144x^3f_{xx}\ .
\label{C2}
\end{equation}
The general  solution of (\ref{C2}) is given by
\begin {equation}
f(x)=C_2x^2+C_1x(144x-1)e^{\frac{1}{144x}}+v_1(x)\ ,
\label{Sol1}
\end{equation}
where 
\begin{equation}
v_1(x)=-864x\left[\frac{1}{144x}+x\ln{x}+\left(\frac{1}{144x}-x\right)Ei\left(1,\frac{1}{144x}\right)e^{\frac{1}{144x}}\right]\ 
.
\end{equation}
Here  
\begin {equation}
Ei(a,z)=z^{a-1}\Gamma(1-a,z)=\int_{1}^{\infty}e^{-zs}s^{-a}ds\ .
\end{equation}
The two integrals of motion are
\begin {equation}
I_1=\frac{144}{x}\left[x(1-144x)(f_x-v_{1x})+\left(288x-2+\frac{1}{144x}\right)(f-v_1)\right]\,,
\end{equation}
\[
I_2=\frac{1}{x(144x-1)e^{\frac{1}{144x}}}\left\{f-v_1-144[x(1-144x)(f_x-v_{1x})+  \right.
\]
\begin {equation}
\left. \left(288x-2+\frac{1}{144x}\right)(f-v_1)]\right\}\ .
\end{equation}

\item Let $\Lambda\neq 0$. Then Eq. (\ref{C1}) has a complex solution, 
which has no physical meaning as it gives a complex action.  
\end{itemize}
Hence, it appears that the $\Lambda$CDM model (\ref{LambdaCDM}) can 
not be reproduced by $R+f(G)$ gravity in the absence of matter. The 
only solution found restricts the Hubble rate to give a decelerated 
Universe.

\subsection{Case II: $\rho_m\neq 0$ and $\Lambda = 0$}

We now explore the case when some kind of matter with a particular 
EoS is present in the Universe, but with no cosmological constant 
term in the Hubble parameter described in (\ref{LambdaCDM}). We 
explore several examples where different kind of matter 
contributions are considered.
 
\subsubsection{Example 1} 
Let us now consider the case when $\Lambda=0$ and the evolution of 
the matter density behaves as
\begin {equation}
\rho_m=3H^2=3x.
\end{equation}
In this case the modified Friedmann equation (\ref{1.3}) reads as
\begin {equation}
0=2f+x(144x-1)f_x-144x^3f_{xx}.
\label{C5}
\end{equation} This equation has two integrals of motion
\begin {equation}
I_1=\frac{144}{x}\left[x(1-144x)f_x+\left(288x-2+\frac{1}{144x}\right)f\right],
\end{equation} 
\begin {equation}
I_2=\frac{1}{x(144x-1)e^{\frac{1}{144x}}}\left\{f-144\left[x(1-144x)f_x+\left(288x-2+\frac{1}{144x}\right)f\right]\right\}.
\end{equation} 
The  general solution of the equation (\ref{C5}) is given by
\begin {equation}
f(x)=C_1x^2+C_2x(144x-1)e^{\frac{1}{144x}} .
\end{equation}
This function reproduces the solution (\ref{LambdaCDM}) under the conditions imposed above.

\subsubsection{Example 2} 
Now  we consider a more general case, where the energy density is given by,
\begin {equation}
\rho_m=u(x)\ ,
\end{equation}
where $u(x)$ is some function of $x$. 
In this case the modified Friedmann equation (\ref{1.3}) reads as
\begin {equation}
0=2[3x-u(x)+f]+x(144x-1)f_x-144x^3f_{xx}\ .
\end{equation}
Its  general solution is
\begin {equation}
f(x)=C_1x^2+C_2x(144x-1)e^{\frac{1}{144x}}+v_2(x)\ , \label{C5b}
\end{equation}
with
\begin {equation}
v_2(x)=288x\left[\left(x-\frac{1}{144}\right)e^{\frac{1}{144x}}J_1-\frac{x}{144}J_2\right]\ ,
\end{equation}
where
\begin {equation}
J_1=\int\frac{3x-u}{x^2}e^{\frac{1}{144x}}dx, \quad J_2=\int\frac{(3x-u)(144x-1)}{x^3}dx\ .
\label{C6}
\end{equation}
Note that the two integrals of motion  are
\begin {equation}
I_1=\frac{144}{x}[x(1-144x)(f_x-v_{2x})+(288x-2+\frac{1}{144x})(f-v_2)]\ ,
\end{equation}
\[
I_2=\frac{1}{x(144x-1)e^{\frac{1}{144x}}}\left\{f-v_2-144\left[x(1-144x)(f_x-v_{2x})+ \right. \right.
\]
\begin {equation}
\left.\left. \left(288x-2+\frac{1}{144x}\right)(f-v_2)\right]\right\}\ 
.
\end{equation} 
In fact, we can directly verify that
\begin {equation}
I_{1x}=I_{2x}=0\ .
\end{equation}
Then, the solution (\ref{C5b}) gives the function of the 
Gauss-Bonnet invariant that reproduces this model for any kind of EoS matter fluid.

\subsection{Case III: $\rho_m\neq 0$ and $\Lambda\neq 0$}
Let us now explore the most general case for the solution 
(\ref{LambdaCDM}) in $R+f(G)$ gravity with a non vanishing matter 
fluid with a given EoS parameter.

\subsubsection{Example 1}

We  consider 
\begin{equation}
\rho_m=3x+\beta\,. 
\end{equation}
Then Eq. (\ref{1.3}) takes the form
\begin {equation}
0=(\beta-f)(\Lambda-2x)+[48x^2(3x-\Lambda)+x(\Lambda-x)]f_x+24x^2(3x-\Lambda)(\Lambda-2x)f_{xx}
\end{equation}
and has the following particular solution:
\begin{equation}
 f(x)=\gamma x^2-\gamma\Lambda x+\beta\,.
\end{equation}
If $\Lambda=0$, then $ f(x)=\gamma x^2+\beta$. Also if $\gamma=0$, 
then the solution takes the form $f=\beta$, which corresponds to the 
cosmological constant. Note that if $\beta=-\Lambda$ then
\begin{equation}
\rho_m=\frac{\rho_{03}}{a^3}=3x-\Lambda, \quad f(x)=\gamma x^2-\gamma\Lambda x-\Lambda\ .
\end{equation}
This gives a solution where the cosmological constant is corrected 
by the contribution from $f(G)$, what may resolve the cosmological 
constant problem.
\subsubsection{Example 2} 

Now  we consider the density of the energy which is given by
\begin{equation}
\rho_m=kx^2+3x+\beta.
\end{equation}
The corresponding first Friedmann equation reads as
\begin {equation}
0=(kx^2+\beta-f)(\Lambda-2x)+[48x^2(3x-\Lambda)+x(\Lambda-x)]f_x+24x^2(3x-\Lambda)(\Lambda-2x)f_{xx}.
\end{equation}
This equation has the following particular solution: 
\begin{equation}
f(x)=\gamma x^2+\left(\frac{k}{72}-\gamma \Lambda\right) x+\beta,
\end{equation}
where we must have $\Lambda=-\frac{1}{24}$. As well as in the above 
example, the given function $f(G)$ produces a relaxation on the 
cosmological constant, which can be seen as the possible resolution 
of the cc problem.

\section{Cosmology in $R+f(G)$ gravity with the presence of an inhomogeneous fluid}

Let us now consider the theory described by the action (\ref{1.1}) in the presence of a perfect fluid, whose EoS is given by the general expression:
\begin{equation}
p=w(a)\rho+\eta(a)\ ,
\label{D1}
\end{equation}
where $w(a)$ and $\eta(a)$ are arbitrary functions of the scale 
factor. This kind of EoS could correspond to a dynamical viscous 
fluid or possibly the effective EoS that accounts for the extra 
terms in the gravitational action, as curvature terms or scalar and 
vector fields (see Ref. \cite{DSG}). Then, the FRW equations 
(\ref{1.3}) are written now as
\begin{equation}
3H^2=\rho+\rho_{f(G)}\ , \quad 2\dot{H}+3H^2=-(p+p_{f(G)})\,.
\label{D2}
  \end{equation}
Here the energy and pressure densities $\rho_{f(G)}$ and $p_{f(G)}$ 
are properly defined to account for the extra terms introduced by 
the $f(G)$ function in the action (\ref{1.1}), and are given by
\[
\rho_{f(G)}=Gf_G-f-24\dot{G}H^3f_{GG}\ , 
\]
\begin{equation}
p_{f(G)}=8H^2\ddot{f_G}+16H(\dot{H}+H^2)\dot{f_G}+f-Gf_G\ .
\label{D3}
\end{equation}  
Then, by combining the FRW equations (\ref{D2}) and using the EoS 
(\ref{D1}), we could write the inhomogeneous term $\eta(a)$ as a 
function of the Hubble parameter:
\begin{equation}
\eta(a)= -w(a)\rho_{f(G)}+p_{f(G)}+2\dot{H}+3H^2(1+w(a))\ .
\label{D4}   
\end{equation}  
Hence, by specifying a cosmic solution $H(a)$, any cosmological 
history can be reconstructed in this framework, where the 
contributions from $f(G)$ and the inhomogeneous fluid described by 
(\ref{D1}) drives the Universe evolution. To show this, let us 
consider the example
\begin{equation}
3H^2=H_0^2+H_1a^m\ ,  
\label{D5}
\end{equation}
where $H_0$, $H_1$ and $m$ are constants. This solution reproduces a 
Universe dominated by an effective cosmological constant at the 
current time and which enters a phantom epoch in the future (as some 
observations suggest). We could consider $w(a)=0$ and the $f(G)$ 
function calculated in (\ref{Sol1}), which reproduces an effective 
cosmological constant. Then, the effective fluid (\ref{D1}) 
describes dust-matter at the beginning, when the cosmological 
constant dominates and then it drives the Universe into a phantom 
epoch, which probably finishes at the so-called Big Rip singularity. 

\section{Cosmological solutions in pure $f(G)$ gravity}

We have studied so far a theory described by the action (\ref{1.1}), 
which is given by the usual Hilbert-Einstein term plus a function of 
the Gauss-Bonnet invariant, that is assumed to become important in 
the dark energy epoch. In this section, we are interested to 
investigated some important cosmic solutions in the frame of a 
theory described only by the Gauss-Bonnet invariant, and whose 
action is given by
\begin{equation}
S=\int d^4x\sqrt{-g}\left[f(G)+L_m\right]\ .
\label{D6}
\end{equation}       
In this case, the  FRW equations are:
\[
0=Gf_G-f-24\dot{G}H^3f_{GG}+\rho_m
\]
\begin{equation}
0=8H^2\ddot{f_G}+16H(\dot{H}+H^2)\dot{f_G}+f-Gf_G+p_m\ .
\label{D7}
\end{equation}
We are interested to explore some important solutions from the 
cosmological point of view, as de Sitter and power law expansions.

\subsection{De Sitter solutions}

De Sitter solutions are described by an exponential expansion of the 
Universe, where the Hubble parameter and the scale factor are given 
by
\begin{equation}
H(t)=H_0\rightarrow a(t)=e^{H_0 t}\ ,
\label{D8}
\end{equation}
where $H_0$ is a constant. This kind of solutions are very 
important, as the observations suggest that the expansion of our 
Universe behaves approximately as de Sitter. It has been shown in 
Ref. \cite{f(R)deSitter} that de Sitter points are critical points 
in $f(R)$ gravity. It is straightforward to see that this is also 
the case in $f(G)$ gravity.  We can explore the de Sitter points 
admitted by a general $f(G)$ by introducing the solution (\ref{D8}) 
in the first FRW equation given in (\ref{D7}), which yields
\begin{equation}
0=G_0f_G(G_0)-f(G_0)\ .
\label{D9}
\end{equation} 
Here, $G_0=24H_0^2$ and we have ignored the contribution of matter. 
Then, we have reduced the differential equation (\ref{D7}) to an 
algebraic equation that can be resolved by specifying a function 
$f(G)$. The de Sitter points are given by the positive roots of this 
equation, which could explain not just the late-time accelerated 
epoch but also the inflationary epoch. The stability of these 
solutions has to be studied in order to achieve a grateful exit in 
the case of inflation, and future predictions for the current cosmic 
acceleration.

\subsection{Power law solutions}

We are now interested to explore power law solutions for a theory 
described by the action (\ref{D6}). This kind of solutions are very 
important during the cosmic history as the matter/radiation epochs 
are described by power law expansions, as well as the possible 
phantom epoch, which can be seen as a special type of these 
solutions. Let us start by studying a Hubble parameter given by
\begin{equation}
H(t)=\frac{\alpha}{t}\rightarrow a(t)\sim t^{\alpha}\ ,
\label{D10}
\end{equation}
where we take $\alpha> 1$. Then, by introducing the solution 
(\ref{D10}) into the first FRW equation (\ref{D7}), it yields the 
differential equation
\begin{equation}
0=-f(G)+Gf_G+\frac{4G^2}{\alpha-1}f_{GG}\ , \label{D11}
\end{equation}
where we have neglected any contribution of matter for simplicity. 
The equation (\ref{D11}) is a type of Euler equation, whose solution 
is
\begin{equation}
f(G)=C_1G+C_2G^{\frac{1-\alpha}{4}}\ .
\label{D12}
\end{equation}
Thus, we have shown that power-law solutions of the type (\ref{D10}) 
correspond to actions with powers on the Gauss-Bonnet invariant, in 
a similar way as in $f(R)$ gravity, where power-law solutions 
correspond to an action with powers on the scalar curvature, $R$ 
(see Ref. \cite{Dunbsy1}). 

Let us now explore another kind of power-law solutions, where the 
Universe enters a phantom phase and ends in a Big Rip singularity. 
This general class of Hubble parameters may be written as
\begin{equation}
H(t)=\frac{\alpha}{t_s-t}\ ,
\label{D13}
\end{equation}
where $t_s$ is the so-called Rip time, i.e. the time when the future 
singularity will take place. By inserting the solution (\ref{D13}) 
into the first FRW equation (\ref{D7}), the equation yields
\begin{equation}
0=-f(G)+Gf_G(G)-\frac{4\alpha^2G^2}{1+\alpha}\ , 
\label{D14}
\end{equation}
which is also a Euler equation, whose solution is given by,
\begin{equation}
f(G)=C_1G+C_2G^{\frac{1+\alpha}{4\alpha^2}}\ .
\label{D15}
 \end{equation}
Thus, we have showed that power law solution of the type 
radiation/matter dominated epochs on one side and phantom epochs on 
the other, are well reproduced in pure $f(G)$ gravity, in a similar 
way as it in $f(R)$ gravity.

\section{Conclusions}

We have explored in this paper several cosmological solutions in the 
frame of Gauss-Bonnet gravity, considering specially the case of an 
action composed of the Hilbert-Einstein action plus a function on 
the Gauss-Bonnet invariant. Also pure $f(G)$ gravity has been 
considered, as well as the possibility of the implication of 
inhomogeneous terms in the EoS of a perfect fluid, which could 
contribute together with modified gravity to the late-time 
acceleration. We have shown that the $\Lambda$CDM model can well be 
explained in this kind of theories, which may give an explanation to 
the cosmological constant problem as the modified gravity terms may 
act relaxing the vacuum energy density. Other kinds of solutions in 
$f(G)$ gravity have been reconstructed. It has been shown that 
$f(G)$ gravity could explain the dark energy epoch whatever the 
nature of its EoS, of type quintessence or phantom, and even the 
inflationary phase.  More complex cosmological solutions would 
require numerical analysis, but our analysis of a few simple cases 
has already shown that $f(G)$ gravity accounts for the accelerated 
epochs and may contribute during the radiation/matter dominated 
eras, and it may explain also the dark matter contributions to the 
cosmological evolution, what will be explored in future works. This kind of modified gravity models which  reproduce dark energy and inflation, can be modeled as an inhomogeneous fluid with a dynamical equation of state, what would be distinguished from other models with a static EoS. Even as perturbations in modified gravity behave different than in General Relativity, it could give a signature of the presence of higher order terms in the gravity action, as the Gauss-Bonnet invariant, when structure formation is studied and simulations are performed, what should be explored in the future. 

\begin{acknowledgements}

DSG acknowledges a grant from MICINN (Spain), project FIS2006-02842. The 
support of the Academy of Finland under the Projects No. 121720 and 
127626 is gratefully acknowledged. 

\end{acknowledgements}

\end{document}